\documentclass[reprint]{elsarticle}

\usepackage{graphics}
\usepackage{graphicx}
\usepackage{epsfig}

\usepackage{amssymb}

\usepackage{float}
\usepackage{color}

\begin{document}

\begin{frontmatter}


\title{Title\tnoteref{label1}}

\title{Development of hard x-ray photoelectron SPLEED-based spectrometer applicable for probing of buried magnetic layer valence states}


\author[label1]{<Xeniya Kozina>}
\address[label1]{<Japan Synchrotron Radiation Research Institute, SPring-8, Hyogo, 679-5198, Japan>}

\author[label2]{<Carlos Eduardo Viol Barbosa>}
\author[label2]{<Julie Karel>}
\author[label2]{<Siham Ouardi>}
\address[label2]{<Max-Planck-Institut f\"ur Chemische Physik fester Stoffe, 01187 Dresden, Germany>}

\author[label3]{<Masafumi Yamamoto>}
\address[label3]{<Division of Electronics for Informatics, Hokkaido University, Sapporo 060-0814, Japan>}

\author[label4]{<Keisuke Kobayashi>}
\address[label4]{<Japan Atomic Energy Agency, SPring-8, Hyogo, 679-5148, Japan>}

\author[label5]{<Gerd Sch\"onhense>}
\address[label5]{<Institut f\"ur Physik, Johannes Gutenberg - Universit\"at, 55099 Mainz, Germany>}

\author[label2]{<Gerhard H. Fecher}

\author[label2]{<Claudia Felser>}
\author[label1]{<Eiji Ikenaga>}

\date{\today}

\begin{abstract}

A novel design of high-voltage compatible polarimeter for spin-resolved hard x-ray photoelectron spectroscopy (Spin-HAXPES)
went into operation at beamline BL09XU of SPring-8 in Hyogo, Japan.
The detector is based on the well-established
principle of electron diffraction from a W(001) single-crystal at a scattering energy of 103.5~eV. It’s special feature is that it can be operated
at a high negative bias potential up to 10~kV, necessary to access the HAXPES range. The polarimeter is operated behind a large hemispherical
analyzer (Scienta R-4000). It was optimized for high transmission of the transfer optics. The exit plane of the analyzer contains a delay-line
detector (20~mm dia.) for conventional multichannel intensity spectroscopy simultaneously with single-channel spin analysis. The performance of
the combined setup is demonstrated by the first spin-resolved data for the valence-region of a FeCo functional layer of a tunneling device,
buried beneath 3~nm of oxidic material. The well-structured spin polarization spectrum validates Spin-HAXPES in the valence energy range as
powerful method for bulk electronic structure analysis.
The spin polarization spectrum exhibits a rich structure, originating from clearly discernible transitions in the majority and minority partial spin spectra.

\end{abstract}


\end{frontmatter}


\section{Introduction}
\label{Introduction}

Over the recent decades emerging spintronic technologies have constantly triggered worldwide scientific interest because of
their successful commercial applications in data storage. Major development efforts have been made to the improvement of magnetoresistive
characteristics of tiny multilayer stacks, - known as magnetic tunnel junctions (MTJs), - a key component of spintronic devices.
This allowed to make a big step forward in shrinking a magnetic bit size and significantly enhance data storage capacity~\cite{Parkin2015}.
Tunnelling magnetoresistive characteristics, quantified by the tunnel magnetoresistance (TMR) ratio, define the efficiency of a single
MTJ and depend on the spin-polarisation of the electrons at the Fermi energy $\varepsilon_{\rm F}$ of the functional materials employed in the particular device.

Spin-polarized photoelectron spectroscopy (SP-PES) is a powerful method providing direct information on the spin-dependent electronic
structure of materials. The method in general is based on scattering of the polarized electron beam at a target material.
Electrons with opposite spins have different scattering
cross sections thus defining the difference in their scattered beam intensities~\cite{Kirschner1985},~\cite{Feder1985}.

Owing to a high surface sensitivity, conventional ultraviolet PES and soft-x-PES are excellently suited for the study of thin adsorbate overlayers,
surface states and topologically-protected edge states that also exist only in the topmost few layers. In such cases, the high surface sensitivity
is an advantage and can be exploited in the experiment.

However, the relevance of quantitative evaluation
of materials spin-polarization obtained by SP-PES has been debated~\cite{Dowben2011},~\cite{Osterwalder2012}. While the corrections for the Fermi velocity as well as for the spin
relaxation times can be neglected in SP-PES, which makes it advantageous among other spin polarization measurements methods, the wave vector orientation and existence of surface have to be taken into account and often take over the major role in the results interpretation.
In particular, for highly reactive surfaces, the small inelastic mean free path is a serious obstacle
because surface contamination usually reduces the spin polarization due to spin-dependence scattering at paramagnetic centers, or, even more seriously,
since the chemisorbates change the surface band structure in the topmost layers. Since the degree of contamination depends on time and spin-resolved
spectroscopy is usually time-consuming, the long acquisition time is often prohibitive for spin-resolved studies of such surfaces.

The situation is different in the hard x-ray range, where these obstacles and nonconformities, in measured and around-state spin-polarization values in VUV and soft-X range,
can be effectively eliminated. In hard x-ray photoelectron spectroscopy (HAXPES) true bulk information is provided ruling out critical surface artefacts.
The inelastic mean free path increases to typical values exceeding $10~nm$. Hence, many atomic layers contribute
to the total signal. This means, that the topmost few layers are not dominant in the spectra and the surface signal contributes only less than 5$\%$ to the total signal~\cite{Suga09}.
Moreover, if the surface is covered by a protective coating
or by a stable contamination layer, like, e.g. an oxide, the spin-resolved signal originating from deeper-lying layers will be stable and not time-dependent.
The most promising aspect is indeed to study buried layers, employed in devices and thus requiring non-destructive probing~\cite{Fecher2008}.
Owing to an information depth exceeding 10~nm~\cite{Koba03}, an active layer in a multilayer device stack
can be studied, even if another functional layer is on top. Due to the element specificity of PES, the element of interest can be chosen by the proper
photoelectron energy. The power of this method has been shown exploiting magnetic circular dichroism in angular distribution of photoelectrons (MCDAD) for gaining magnetic information~\cite{Kozina2011},~\cite{Fecher2014}. It can be
expected that spin-resolved HAXPES will be applied to many technologically relevant samples, once the detection technology has reached a sufficient maturity.
In very few experiments, it was possible to measure the spin polarization spectrum of core levels of buried layers~\cite{Gloskovskii2012},~\cite{Stryganyuk2012},~\cite{Ueda2014}.

Spin-resolved photoemission measurements in the hard x-ray range are extremely promising in
regard of the outcome from their application to spintronic-device characterization and thereby future data storage technologies development as well as for fundamental research.

In the present paper, we show the first spin-HAXPES implementation and development results for the valence range taken at the beamline BL09XU of Spring-8 in Japan.
This analysis system employing spin detector based on SPLEED at W(001) demonstrates for the first time that the electronic structure close to the Fermi energy $\epsilon_{\rm F}$
and the spin polarization of deeply buried magnetic layers imbedded as electrodes in spintronic devices can be measured.



\section{Experimental setup}
\label{Experimental setup}

Implementation of spin-HAXPES is unavoidably related to significant losses in signal intensity (at the spin discriminating system and reduced
cross sections of high-energy photons~\cite{Fecher2007}). This poses particular limitations and requirements to the spin-detection approach and comes along with instrumental development.
The high energy of photoelectrons ejected from a sample assumes large retardation ratios which leads to aberrations and large angular divergence
of the electron beam at the target, thus decreasing spin sensitivity and total transmission.

Spin detection approaches, based on either spin-orbit or exchange
interaction, include high-energy Mott scattering~\cite{Sherman56}, very low electron diffraction (VLEED)~\cite{Okuda2008}, and spin-polarised low energy electron
diffraction (SPLEED)~\cite{Sawler1991},~\cite{Yu2007}. Spin-detection efficiency depends on its spin-sensitivity or Sherman function~\cite{Sherman56},~\cite{Kessler1985}, - intrinsic to
a particular detector configuration. Mott-type spin polarimeters~\cite{Kisker82}, employing back-scattering of electrons at high-Z polycrystalline targets
became the most commonly used due to their simplicity in target treatment, the inert target, and, consequently,
stable operation. Nevertheless, high operation voltages (tens of keV), and the low Sherman function S of this detector type (0.1-0.2), and, correspondingly,
the difficulties in spin-resolved valence states detection make it less advantageous in comparison with detectors based on diffraction
at a single crystalline target, which consequently leads to enhanced spin sensitivity and larger signal-to-noise ratio. Among those VLEED detectors, where spin selection occurs due
to exchange-scattering at a Fe$(001)$-p$(1\textrm{x}1)$-O target exhibit a high Sherman function (S=0.4) and thus figure of merit (FOM) in the order of {10$^{-2}$}. In contrast to
Mott type detectors, the scattering energy in VLEED detectors is very low in the range of several eV (6-12~eV). First, developed by
Kirschner and Feder in 1979~\cite{Kirschner1979}, and later widely used~\cite{Sawler1991} and commercialized~\cite{Yu2007}, SPLEED type spin polarimeters, using (2,0)
diffraction reflexes from a W(001) surface, occupy a middle position between the above mentioned
detector types  with S=0.25, FOM$\approx10^{-4}$ and 104.5~eV scattering energy. This type of detector became very common and is
successfully used for spin-resolved valence states detection of materials.
Owing to its reasonable spin-sensitivity value
and at the same time relatively high scattering energy (fitting well to typical pass energies of analyzers)
the SPLEED detector is more compatible among the above
mentioned single-channel detector types to accomplish spin-HAXPES implementation goals.
Recently a 2D-multichannel approach has been established which significantly enhances the FOM by 4 orders of magnitude~\cite{Kolbe2011}.
So far this method has been limited
to the use of VUV energy range only, and the development of a high-energy range version of this type of detector is currently in progress
and therefore is not considered here within the present research frames.

\begin{figure}[htb]
\centering
   \includegraphics[width=9cm]{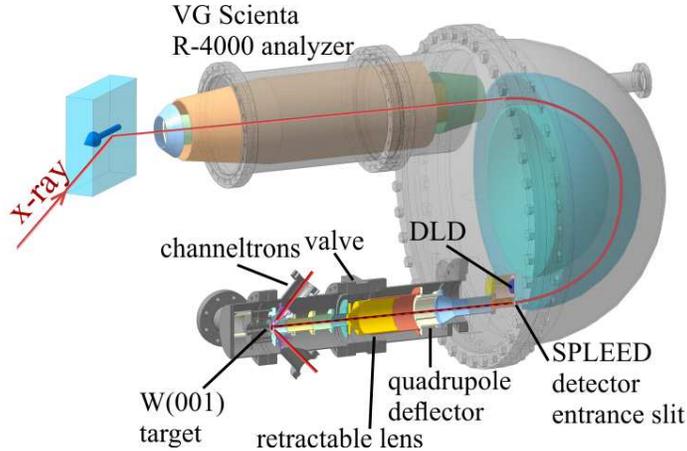}
   \caption{Schematic layout of the spin-HAXPES experimental setup, employing VG Scienta high-energy analyser and SPLEED detector based on W(001)}
\label{fig:setup}
\end{figure}

The layout of the spin-HAXPES system experimental setup is shown schematically in Figure~\ref{fig:setup}.
The SPLEED polarimeter is attached to a high-energy VG Scienta R4000 hemispherical high-energy analyzer,
substituting a common MCP-Screen-CCD camera-detector at its exit port.


In the modified exit port we implemented (i) a small delay-line detector (DLD) (20~mm diameter) for multichannel intensity detection
which is placed off-center, and (ii) the entrance of the spin detector optics, consisting of a slit with size 2$\times$12~mm$^{2}$. This
combination allows to perform normal intensity spectroscopy with less than a factor of 3 reduction in efficiency
and simultaneously single-channel spin polarimetry. The latter is facilitated by focusing the energy-selected part of the spectrum falling through the 2$\times$12~mm$^{2}$
discrete exit slit into the transfer optics of the spin detector. Owing to its 12~mm length in non-dispersive direction, the spin branch
integrates over a large angular range, which is no disadvantage for the present experiment in the hard x-ray range. Given the pass
energy of 200~eV, the 2~mm slit width selects an energy band of 1000~meV from the spectrum.
The center of the DLD is displaced
with respect to the analyzer exit port axis by 8~mm towards the inner hemisphere.

The intensity losses, related to implementation of spin-resolved methods into hard x-ray range, are accounted for the losses due to the
electron focusing system throughput including both analyzer and SPLEED detector,
which depends on the interplay between electron beam trajectories at the exit of the analyzer and at the entrance of the SPLEED detector.
The beam deflection in the dispersive plane, intrinsically for this analyzer type, leads to the presence of a certain angular divergence of the
electron trajectories at the exit plane in energy dispersive direction.
SPLEED detection requires
minimal possible electron beam angular divergence at the entrance plane. To overcome this drawback in the present system the priority is given
to achieving the most preferable conditions for the electron
focusing system of the spin-detector. Therefore the axis of the SPLEED detector electron transfer optics is displaced from the analyzer
exit port axis towards the outer hemisphere by 24~mm. In this point the exit rays run perpendicular to the exit plane.
However, the single-channel solution reduces to a large extent the spatial acceptance of
the spin-detector entrance slit, which is only 1.5\% of the total acceptance of a common MCP-CCD camera detector.



High-energy photoelectrons, ejected from a magnetically ordered sample, are accepted by the entrance lens and focused at the analyzer
entrance slit (typically 0.5~mm or 1.5~mm vertical size). After being discriminated by their energies at the exit port, the electrons pass
through the slit aperture of 2$\times$12~mm$^{2}$ size and
are fed into the spin-detector focusing system and decelerated from the analyzer pass energy of 200~eV down to the 103.5~eV
scattering energy of the W(001) target crystal, which corresponds to the optimum FOM of a SPLEED type detector. To provide the
maximum possible resolution simultaneously with the maximal possible countrate, the entrance slit vertical size of the hemispherical analyzer
and the spin-detector are chosen to be similar in size (1.5 mm and 2 mm correspondingly) at an analyzer pass energy of 200~eV.
The analyzer entrance slit size of 25~mm (horizontal(H))$\times$1.5~mm (vertical(V)) and magnification of Scienta R4000 objective lens equal to 5 give 5~mm (H)$\times$0.3~mm (V)
observation area at the sample position. The photon beam footprint at the sample is equal to 706 (H) x 4.6 (V)~$\mu$m$^{2}$, being much smaller than the analyzer
observation area. This assures the full acceptance and transmittance of the photoelectrons and thus minimum intensity losses at this stage.

The spin-detector electron focusing system is designed to be compatible for the high-energy operation together with the Scienta R-4000 analyzer, thus being floating at a high negative potential,
and consists of two stages. The spin-detector lens elements are supplied independently with respect to the analyzer Herzog potential. The front stage provides the maximal beam transmission and the rear one with its Einzel lens part
facilitates the final focusing correction close to the target surface. Off-axial beam divergence correction is performed at
the front stage by a set of quadrupole deflector optics.

Extensive simulations of electron trajectories were performed, by means of the SimIon software, in order to optimize the transmission of the transfer optics. The main obstacle is the fact that
the angular divergence in dispersive direction is $\pm$3$^{\circ}$ and the energy at the exit is approximately 2 times larger than the scattering
energy at the spin detector crystal. Beam shape in such a system has to obey Liouville’s theorem:

\begin{equation}
\textrm{M} \sin \alpha \sqrt{E} = \emph{const}
\label{eq:Liouville}
\end{equation}

with M, $\alpha$ and $E$ denoting the lateral magnification, the angular divergence and the beam energy, respectively. A deceleration of the
beam thus necessarily increases the lateral size and/or the angular divergence of the beam. Another obstacle is the shape of the exit
slit which induces a strong astigmatism in the imaging optics. The focusing size of 5.6~mm at the target crystal plane with the [$\pm$8$^{\circ}$]
beam angular divergence is achieved at the pass energy of 200~eV and
corresponds to 97.5\% transmission of the transport lens between exit plane and SPLEED crystal.
Moreover, due to the frequent preparation of the spin-detector crystal
a closing valve between the main chamber and the spin detector chamber is indispensable. Figure~\ref{fig:setup} shows that
the valve is implemented
about midway between analyzer exit plane and spin detector crystal, which introduces 8~mm potential-free space in between neighbouring lens elements.
Since the whole optics is floating on a high negative
potential (typically 6 to 10~kV) in HAXPES experiments, the grounded walls of the valve have to be effectively shielded in the lens
system. For this purpose we introduced a movable lens element (RL) that is shifted inside the valve body during measurements.

The spin-discriminated signal obtained at the W(001) clean single crystalline surface in the (2,0) series of low energy electron diffraction maxima
is detected by two pairs of channeltron detectors. High resistive channeltrons, geometrically set in the back scattering sphere surface of the target
crystal, provide less than 2~cps of background noise which assures their compatibility, due to sufficient detection sensitivity, for such an extremely
low countrate experiment. The channeltrons are equipped with retardation grid which repels secondary electrons.






\section{Results and experimental procedures}
\label{Experimental setup}

As it is widely described in literature, the spin sensitivity obtained by diffraction at W(001) is dependent on the target surface condition,
and thus is a decaying function of time. In UHV the lifetime of the W(001) surface is in the order of an hour and requires periodic treatment
procedures~\cite{Zakeri2010}. The base pressure in the present spin-detector chamber was 3.0$\times$10$^{{-10}}$ mbar and thus the lifetime of the clean crystal
surface is evaluated to be about 30~min without CO-flash procedures. As depicted in Figure~\ref{fig:setup} the SPLEED detector chamber is supplied by a
valve which separates the detector from the main chamber, and allows target flashing to remove CO and H contamination, without interfering the vacuum in the detector
front part and the spectrometer itself. In the rear part of the spin-detector chamber the oxygen doser is installed. A silver tube heated by an inner thermocoax heater
provides precise dosing of high purity oxygen.
The crystal treatment procedures are performed by means of electron bombardment.
The heating unit, installed behind the crystal, consists of a filament and Wehnelt to
avoid heat dissipation as well as to enable homogeneous, precise and short crystal flashes.
The preparation procedure parameters were optimized in a separate UHV setup equipped by LEED as well as a mass spectrometer.
The target was treated for a long time before the measurements, tracing the W(001) surface quality by LEED, and afterwards the target was installed in spin detector chamber for the measurements.
The target surface preparation was made in 3 stages starting from "hot oxygen treatment" heating at 1473~K in oxygen atmosphere (5$\times$10$^{{-8}}$~mbar) for 20~min, followed by room temperature (RT) passivation
at the same oxygen partial pressure, and finally a high temperature flash at 2473~K for 6~s at the base pressure.
During the measurements the W(001) treatment was made repeatedly after every 30~min of measurement time.

\begin{figure}[htb]
\centering
   \includegraphics[width=8cm]{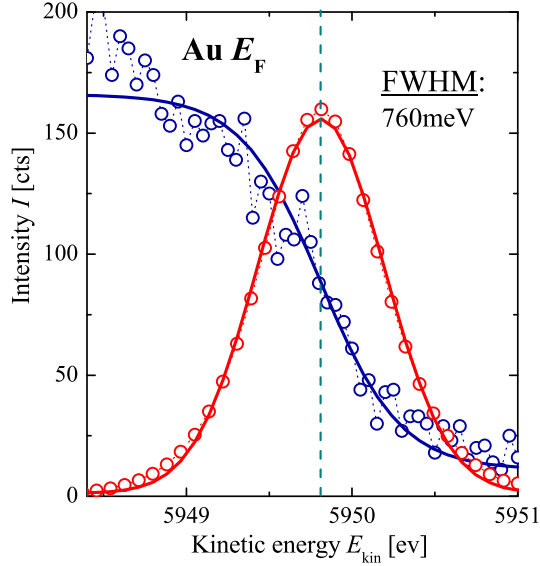}
   \caption{The Fermi edge spectrum obtained at 5.95~keV excitation energy from an amorphous Au film.
            The measurements were performed at an analyser pass energy of 200~eV at room temperature. }
\label{fig:AuEf}
\end{figure}

The spin-HAXPES implementation was made at the hard x-ray undulator beamline BL09XU at SPring-8, Hyogo, Japan.
High photon flux provided by this 3rd generation synchrotron facility makes it an ideal and almost unique place to perform such challenging experiments suffering from very low countrates.
A photon energy of 5.95~keV chosen for this experiment assures the necessary bulk sensitivity and still sufficiently high ionization cross section which drastically decreases at higher excitation energies.
The beam bandwidth of 50~meV, which delivers $\approx 10{^{11}}$~photons/s photon flux at the focus position, is obtained by the primary Si(111) double crystal monochromator and
the following downstream Si channel-cut monochromator where the (333) reflection plane is used~\cite{Ikenaga2013}.
The beam, initially focused by a standard vertical mirror setup, is focused down to 12.4 (H) x 4.6 (V)~$\mu$m$^{2}$ by a pair of Kirkpatrick-Baez mirrors.
The experiment was performed in near normal emission geometry with 90$^{\circ}$ fixed angle between energy analyzer objective lens axis and photon incidence direction, and a take-off angle of 88$^{\circ}$ was used.
The total energy resolution (analyzer and spin-detector) of 760~meV was verified by Fermi cut-off measurements of an amorphous Au sample at RT (Figure~\ref{fig:AuEf}), 200~eV pass energy  and 1.5~mm analyzer entrance slit size. As the total energy resolution is defined by the instrumental resolution, photon beam resolution (50~meV) and thermal broadening at RT ($\sim$120~meV). Hence the instrumental resolution of 749~meV
was derived by deconvoluting the other contributions from the obtained spectrum.

\begin{figure}
\centering
   \includegraphics[width=6cm]{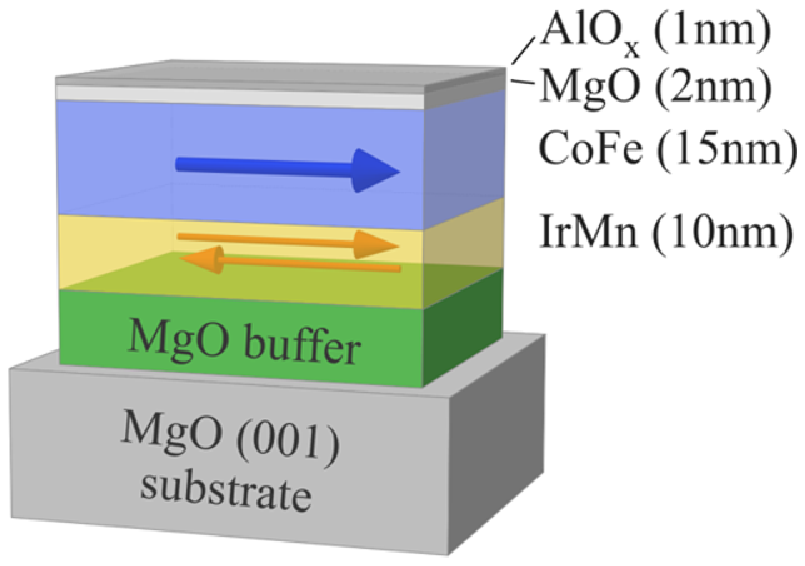}
   \caption{Schematic representation of the exchange-biased multilayer stacks. The structure corresponds to the lower part of an MTJ employing a CoFe functional layer.
            The arrows schematically depict the exchange-biasing configuration used: the ferromagnetic CoFe layer is biased by the bottom-lying IrMn antiferromagnetic layer.}
\label{fig:sample}
\end{figure}

The spin-resolved measurements were performed on magnetized samples mounted on a motorized high-precision $XYZ\theta$ stage and were analyzed under the base pressure of
2.0$\times$10$^{{-9}}$~mbar. In the present experiment exchange-biased multilayer structures with epitaxial CoFe functional layer were used for the measurements.
The structures corresponding to the lower electrode of an MTJ were deposited as follows:
MgO(100) substrate~/ MgO buffer (10\,nm)~/ Ir$_{78}$Mn$_{22}$ (10\,nm)~/ CoFe (15\,nm)~/ MgO barrier (2\,nm)~/ AlO$_x$ (1\,nm)~\cite{Masuda08} (see Figure~\ref{fig:sample}).
To provide fixed exchange biasing of the CoFe layer film
through the IrMn/CoFe interface the deposition was followed by annealing at $350^{\circ}$C for 1~h in vacuum of $5\times 10^{-4}$~mbar in a
magnetic field of 0.4~MAm$^{-1}$~\cite{Marukame07}. An AlO$_x$ layer is used to protect the hygroscopic MgO barrier and thus the stack in whole from any possible contaminations.
An advantage of HAXPES is its bulk sensitivity which enables simultaneous detection of signals from different layers in a multilayer stack.
However, in particular cases (i.e. same chemical elements contained in different layers) this is undesirable and leads to overlapping signals.
The latter becomes unavoidable especially in the valence range. Therefore in the present case a 15-nm-thick buried CoFe layer was grown epitaxially to assure that any photoemission signal
from deeper layers is effectively eliminated.

A pair of identical samples with opposite magnetization was fixed mounted onto a sample holder to assure fast and accurate sample adjustment procedure during measurements.
High precision alignment was made on the samples orientation and tilt with respect to each other and the sample holder plate respectively. The measurements were performed
with $p$-polarized x-rays and the sample magnetization orientation parallel and antiparallel to the beam propagation which corresponds to non-chiral geometry.
Moreover, with the aim to exclude a possible instrumental asymmetry in the spin-resolved photoemission experiment spectra were taken
from both samples. Assigning $I^{+}$ to the signal intensity obtained from one of those samples and $I^{-}$ to the other one, the spin-polarization
was derived as:

\begin{eqnarray}
    P& =       &\frac{1}{S}\frac{\sqrt{I^{+}_{1}I^{-}_{2}}-\sqrt{I^{+}_{2}I^{-}_{1}}}{\sqrt{I^{+}_{1}I^{-}_{2}}+\sqrt{I^{+}_{2}I^{-}_{1}}},
  \label{eq:SP_instrum},
\end{eqnarray}

where 1 and 2 subscripts assign the signals detected by the channeltron detectors 1 and 2, respectively, and \emph{S} is the effective Sherman function, for this detector type, equal to 0.25.
Further, by taking into account the spin-integrated signal $\langle{I}\rangle$

\begin{eqnarray}
    \langle{I}\rangle& =       & \frac{I^{+}_{1}+I^{-}_{1}+I^{+}_{2}+I^{-}_{2}}{4}
  \label{eq:totalI},
\end{eqnarray}

the energy dependent spin-resolved intensities were derived as:

\begin{eqnarray}
    I^{\uparrow}& =       & \langle{I}\rangle(1+P)
  \label{eq:spin-resolved_I_up},
\end{eqnarray}

\begin{eqnarray}
    I^{\downarrow}& =       & \langle{I}\rangle(1-P)
  \label{eq:spin-resolved_I_down},
\end{eqnarray}

\begin{figure}[htb]
\centering
   \includegraphics[width=8cm]{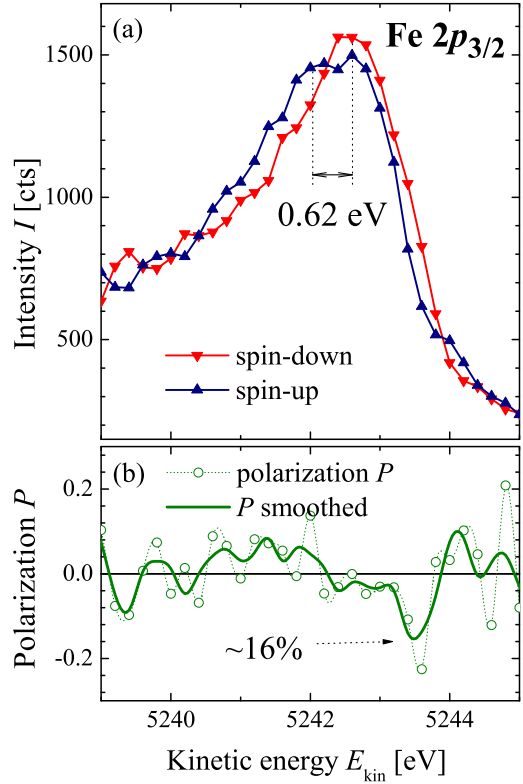}
   \caption{Spin-resolved HAXPES Fe$2p_{3/2}$ spectra from remanently magnetized CoFe deeply buried layer.
            Spin-up (majority) and spin-down (minority) data as well as spin-polarization are derived using equations~\ref{eq:spin-resolved_I_up},~\ref{eq:spin-resolved_I_down} and~\ref{eq:SP_instrum} respectively.}
\label{fig:Fe2p}
\end{figure}

Figure~\ref{fig:Fe2p}(a) shows the spin-resolved spectra in the range of the Fe $2p_{3/2}$ core-state.
The data were accumulated for 36~min in total for both samples.
A significant difference is evident between the spin-up and spin-down energy distribution curves plotted in blue (up triangles) and red (down triangles)
respectively.
As it is seen the exchange interaction shifts the spin-up spectrum towards higher binding energy and a broadened
majority signal appears at 5242~eV. In contrast, the
spin-down spectrum exhibits a narrower and intensity enhanced peak with its maximum at 5242.6~eV.
These two spectral features are assigned to the pure spin states, with m$_{j}$ = $\pm{3/2}$ separated by exchange interraction of the spin-polarized valence band of ferromagnetic
CoFe and the $2p$ core-hole. The exchange splitting is estimated to be 0.62~eV which is in good agreement with previously reported values for a CoFe
buried layer evaluated from MCDAD measurements~\cite{Kozina2011}. The slight quantitative discrepancies are due to different experimental conditions of both experiments, i.e. limitations imposed by instrumental resolution.

The energy dependence of the derived spin-polarization $P$ of the Fe $2p_{3/2}$ state is presented in Figure~\ref{fig:Fe2p}(b).
Having considered the number of data points being possibly resolved within the instrumental resolution,
equal-weight 10-point-average smoothing was applied to the spin polarization data.
The curve exhibits a $"plus-minus"$ feature in this energy range and the evaluated spin-polarization values are +4$\%$ and -16$\%$ as retained for the deep Fe$2p_{3/2}$ core-level
electrons passing through the capping overlayers.
This agrees well with the theoretical calculations made within one-electron theory~\cite{Menchero1998} and demonstrates the result comparable with those reported for FeNi~\cite{Ueda2014} and CoFe buried layers~\cite{Fetzer2015},~\cite{Fecher2014}.

\begin{figure}[htb]
\centering
   \includegraphics[width=8cm]{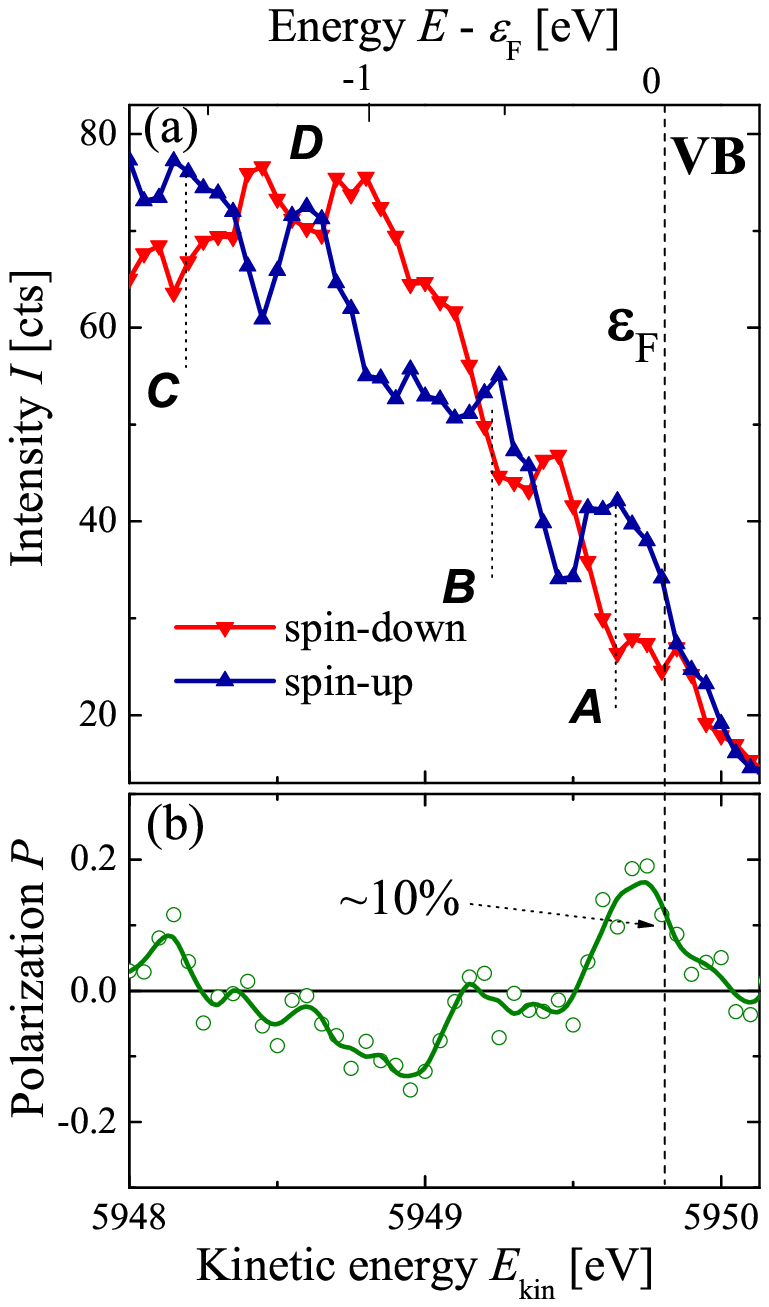}
   \caption{Spin-resolved HAXPES valence band spectra from remanently magnetized CoFe deeply buried layer.
            Spin-up (majority) and spin-down (minority) partial spectra (a) as well as spin-polarization (b) are derived using equations~\ref{eq:spin-resolved_I_up},~\ref{eq:spin-resolved_I_down} and~\ref{eq:SP_instrum} respectively.
            $A$ - $D$ assign peaks referred to in the text. The vertical dashed line assigns the Fermi edge confirmed by measurements on a Au sample.
            The measurements are taken at 5.95~keV excitation energy at RT.}
\label{fig:Ef}
\end{figure}

The first results obtained with the present experimental setup in the range of the valence states close to $\varepsilon_{\rm F}$ are shown in Figure~\ref{fig:Ef}.
Panel (a) displays the partial majority- and minority-spin photoemission spectra as up triangles (blue curve) and down triangles (red curve), respectively.
Panel (b) shows the corresponding spin-polarization spectrum. These data were taken
from a remanently-magnetized single crystalline $bcc$-CoFe~(001) film buried under 3~nm of oxidic material as sketched in Figure~\ref{fig:sample}.
The spectra were accumulated for 170~min in total for both samples. This resulted in 24~h measurement time on the whole including the magnetization-reversal procedure
and the frequent W(001) crystal preparation cycles.
The two spin channels exhibit significantly different spectral features. The majority spectrum exhibits a steep rise at the Fermi edge and a pronounced peak ($A$) at 200~meV below $\varepsilon_{\rm F}$.

This strong majority signal shows up as a pronounced maximum of about 20\% in the spin-polarization spectrum, panel~(b).
The width of the edge in the majority spectrum reflects the total instrumental resolution and this gives a measure for the significance of the observed features.
At 600~meV the broader feature $B$ occurs that is accompanied by almost vanishing spin polarization. On the left-hand side at 1.6~eV the maximum $C$ is visible,
characterized by a positive polarization value close to 10\%. The minority-spin spectrum shows a delayed onset at $\varepsilon_{\rm F}$ but then rises steeply
to a pronounced maximum D at about 1.2~eV. The minority character of this feature is clearly revealed by the pronounced negative spin polarization of –15\%.


The results are consistent with theoretical calculations of the bulk DOS for $bcc$ CoFe alloys~\cite{Belhadji2011},~\cite{Zhang2004} as well as with the experimental results obtained for
FeCo(001)-based MTJs at low excitation energies~\cite{Bonell2012}.
The features $C$ and $B$ of the majority channel are attributed to the high-intensity maximum of theoretical $\Delta_{\rm 1}{\uparrow}$ and lower-intensity $\Delta_{\rm 5}{\uparrow}$ bands of CoFe.
The $\Delta_{\rm 5}{\downarrow}$ bands contribute as the main maximum of the minority spin subbands obtained experimentally (peak $D$).
The spectral weight right at $\varepsilon_{\rm F}$ is mainly dominated by majority states since the high-energy shoulder of the peak $A$ crosses the Fermi energy.
The down-triangles reveal that still a small amount of minority states emission is present as well.
The latter can be assigned to the flat $\Delta_{\rm 5}{\downarrow}$ minority band crossing $\varepsilon_{\rm F}$
and additional contribution from $\Delta_{\rm 1}{\downarrow}$ subbands located above $\varepsilon_{\rm F}$, but shifted towards higher binding energies and therefore imposed at $\varepsilon_{\rm F}$
when the ratio of Co:Fe in the alloy is 50:50.

The majority states in the vicinity of $\varepsilon_{\rm F}$ are imposed by $\Delta_{\rm 1}{\uparrow}$ bands.
According to Belhaji $et. al.$~\cite{Belhadji2011} the DOS weight of $\Delta_{\rm 1}$ minority channel at $\varepsilon_{\rm F}$ is less than
that of the majority channel.
In addition, although quantitatively, the weight of the $\Delta_{\rm 5}{\downarrow}$ bands at $\varepsilon_{\rm F}$ is around 4 times higher
being compared to that of $\Delta_{\rm 1}{\uparrow}$,
the experiment shows the opposite tendency. Such a discrepancy can be explained by the fact that at HAXPES excitation energies the photoemission cross sections
for $p$-states are reduced by almost an order of magnitude more than those for the $s$-states.
This symmetry-dependent difference in experimental cross sections results in a different DOS redistribution for both spin-channels.

The extrema of the spin polarization spectrum correspond to about -15\% and +20\%.
The spin polarization at $\varepsilon_{\rm F}$ is estimated to be +10$\%$,
though the direct comparison of experimental and theoretically reported results
is not straightforward owing to the above-mentioned cross section reduction.
A major difference between low- and high-energy photoemission is the fact that the overlap of the strongly oscillating final state wavefunction in HAXPES with
initial $p$-orbitals drops much stronger with increasing photon energy than for $s$-orbitals. In intensity spectroscopy this fact can be
utilized for the determination of the spectral contributions for states with different symmetry.

In spin-resolved spectroscopy, however, we have to consider that the spinor functions of mixed states contain prefactors from
the matrix elements (besides Clebsch-Gordan coefficient).
This leads to quantitatively different spin polarization values of photoelectrons from the same initial orbitals in (soft-) XPS and HAXPES.
In the present case this means that the $\Delta_{\rm 5}$ ($spd$) contributions in the spectra are diminished in relation to the $\Delta_{\rm 1}$ ($sp$)
contributions.
Unlike intensity spectroscopy, these different components can have different sign of spin polarization, so they can
tend to cancel one another. In turn, the spinpolarization observed in HAXPES on the one hand differs from spin-resolved soft x-ray spectroscopy data,
and, on the other hand, also from the spin polarization observed in transport experiments.

\section{Conclusions and outlook}
\label{Conclusions and outlook}

A spin-resolved hard x-ray photoelectron spectroscopy (spin-HAXPES) instrument utilizing a special, high-voltage compatible W(001) based SPLEED detector,
installed at BL09XU, SPring-8 has been developed.
The resolving power was confirmed by Fermi energy cut-off measurements of Au and an instrumental resolution of about 700~meV was achieved at 5.98~keV excitation energy.
The new instrument was applied to characterize a $bcc$ CoFe(001) deeply buried magnetic layer embedded into a half-MTJ device.
Magnetic properties of the device were studied by core-level photoemission measurements as demonstrated for the Fe $2p_{3/2}$ state.
The exchange splitting was estimated to be 0.62~eV which is consistent with previously reported results obtained for a similar material.
Furthermore, the developed instrument enabled the first measurements performed in the region of the valence states of buried layers, so far inaccessible for direct non-destructive analysis.
Direct spin-resolved probing of a CoFe single-crystalline-MTJ electrode layer resulted in a value of +10\% of the derived spin-polarization at $\varepsilon_{\rm F}$ and extrema
of +20\% and -15\% associated with the maxima caused by transitions from $\Delta_{\rm 1}{\uparrow}$ and $\Delta_{\rm 5}{\downarrow}$ bands, respectively.
The spin-resolved measurements employing SPLEED did not require modification of sample surface, allow direct non-destructive probing of nanostructured devices and therefore open a
broad range of possible investigations for applications for spintronic devices. However, as it was also shown the instrument is extremely close to the principal limits of its performance.
Optimization of the routine procedures are being made.
The present setup has thus reached the maximum possible detection efficiency for this type of detector.
Further significant efficiency improvement is only possible towards multichannel detection~\cite{Kolbe2011, Jourdan2014}.


\section{Acknowledgments}
\label{Acknowledgments}

This work was funded by the Deutsche Forschungs Gemeinschaft (DFG) and the Japan
Science and Technology Agency (JST) (grant no.: FE633/6-1). The work of the
Mainz group was partially supported by DFG (Project TP 1.3-A of the Research Unit 1464 {\it ASPIMATT}) and BMBF (05K13UM4).
The work at Hokkaido University
was partly supported by a Grant-in-Aid for Scientific Research (A) (Grant No.
20246054) from the MEXT, Japan, and by the Strategic International Cooperative
Program of JST. The synchrotron radiation HAXPES experiments were performed at
Bl47XU and BL09XU with the approval of the Japan Synchrotron Radiation Research Institute
(JASRI) (Long-term Proposal 2014A0043, 2014B0043 and General Proposal 2014A1354).
The authors grately acknowledge Prof. Dr. Hans-Joachim Elmers for the valuable and
fruitful discussions on the present results.


\bibliographystyle{model1a-num-names}
\bibliography{CoFe_spinHAXPES_xk}







\end{document}